# Metal-Insulator Transition and $J_{\text{eff}}$ = 1/2 Spin-Orbit Insulating State in Rutile-based IrO$_2$/TiO$_2$ Superlattices

Xing Ming,[1,2] Kunihiko Yamauchi,[3] Tamio Oguchi,[3] Silvia Picozzi[2]*

1. College of Mathematics and Physics and Hubei Key Laboratory for Processing and Application of Catalytic Materials, Huanggang Normal University, Huanggang 438000, PR China

2. Consiglio Nazionale delle Ricerche CNR-SPIN, UOS L'Aquila, Sede Temporanea di Chieti, 66100 Chieti, Italy

3. ISIR-SANKEN, Osaka University, 8-1 Mihogaoka, Ibaraki, Osaka, 567-0047, Japan

* Email: silvia.picozzi@spin.cnr.it

By combining 5$d$ transition-metal oxides showing pronounced spin-orbit interactions and oxide-based heterostructures, we propose rutile-based IrO$_2$/TiO$_2$ superlattices as promising candidates for unconventional electronic properties. By means of density-functional-theory simulations complemented with Hubbard-like corrections, we focus on the evolution of the electronic structure as a function of the IrO$_2$ layer thickness and predict the heterostructures to exhibit a thickness-controlled metal-to-insulator transition, crucially related to the connectivity among IrO$_6$ octahedra. The subtle interplay between electron correlation and spin-orbit coupling leads to an almost pure $J_{\text{eff}}$ = 1/2 spin-orbit insulating state at the level of atomically-thin IrO$_2$ monolayer with almost isolated IrO$_6$ octahedra, leading to a predicted emerging state awaiting for experimental confirmation.

Recently, spin-orbit coupling (SOC) in correlated materials has attracted considerable interests [1, 2]. Indeed, due to several competing energy scales (such as on-site Coulomb interaction $U$, Hund's coupling $J_H$, SOC, crystal field splitting), 5$d$ Ir oxides constitute an excellent playground to uncover fascinating physical properties [3-5]. For example, the Ruddlesden-Popper series $Sr_{n+1}Ir_nO_{3n+1}$ ($n$ = 1, 2 and ∞) show dimensionality -controlled metal-insulator transition (MIT) and correlated metallic states [6]. A novel $J_{eff}$ = 1/2 Mott insulating state has been discovered in quasi-two-dimensional $Sr_2IrO_4$ ($n$ = 1) due to the interplay between electron correlations and SOC [7, 8]. In contrast, the bilayer $Sr_3Ir_2O_7$ ($n$ = 2) is in close proximity to a MIT, whereas the three dimensional counterpart, $SrIrO_3$ ($n = \infty$), is found to be semimetallic with unusually narrow bandwidths [9], also used as a key building block for interfaces [10, 11]. Rich topological states have been realized, including topological magnetic insulators with quantum anomalous Hall effects, nontrivial valley insulators, topological insulators [12], spin-orbit magnetic insulator [13], and magnetic easy-axis reorientation [14].

Inspired by the emergence of new exotic states and potential applications in spintronics [15], metallic binary Ir oxide ($IrO_2$) has gained increased attention among iridates. Experimentally, $IrO_2$ is a Pauli paramagnet without any magnetic ordering down to low temperature, but exhibits a remarkably large spin-Hall resistivity and a moderately high resistivity even in the metallic state [15]. The role of SOC in the $IrO_2$ electronic structure is still under extensive debate. While Miao *et al*. [16] claimed that SOC was not strong enough to induce a MIT in $IrO_2$, x-ray absorption spectroscopy (XAS) [17] and resonant x-ray diffraction [18] experiments reflected the presence of strong SOC and complex $J_{eff}$ = 1/2 orbital states. Furthermore, via hard x-ray photoelectron spectroscopy and first-principles calculations, $IrO_2$ was suggested to well follow the Goodenough model for conductive rutile oxides, explaining the metallic band structure without any $J_{eff}$ = 1/2 Mott insulating state [19]. In contrast, based on a model Hamiltonian, Ir 5$d$ $t_{2g}$ states were proposed to largely retain the $J_{eff}$ = 1/2 character at the Fermi level ($E_F$) even in metallic $IrO_2$ [20]. Recently, optical conductivity measurements and first-principles calculations showed that SOC should play an important role, although XAS measurements did not confirm the formation of $J_{eff}$ = 1/2 state in metallic $IrO_2$ [21].

We recall that $IrO_6$ octahedron is the common crystal basis block in $IrO_2$ and other iridates,

due to the combination of SOC and large $e_g$-$t_{2g}$ crystal field splitting, the sixfold degenerate Ir $t_{2g}$ states are split into quartet $J_{eff} = 3/2$ and doublet $J_{eff} = 1/2$ states (see inset of Fig. 2 (a) and (b)) [3, 7]. Although the $J_{eff} = 1/2$ state is a common ingredient in iridates, its "purity" is often lowered due to structural distortions, so that $J_{eff} = 1/2$ and $J_{eff} = 3/2$ states are mixed, shifting away from the ideal limit of a half-filled band [2, 3, 19, 22]. In order to obtain the $J_{eff} = 1/2$ state, strategies include searching for nearly isolated octahedra in iridium oxides and fluorides [23-25] or growing interfaces/superlattices with other transition-metal oxides [26]. By combining these approaches, we construct a series of artificial $IrO_2$/$TiO_2$ superlattices. Within the framework of density-functional theory (DFT) (for technical details see Supplemental Material [27]), the electronic structure of the superlattices was tuned by changing the $IrO_2$ thickness $m$, in turn linked to the connectivity of $IrO_6$ octahedra. Taking SOC into account, upon increasing on-site Coulomb interactions within DFT + $U$, we predict a MIT from nonmagnetic (NM) metal to antiferromagnetic (AFM) metal and finally to AFM insulator. With a fixed moderate value of Hubbard $U$, the superlattices exhibit a thickness-controlled MIT, varying from AFM metal ($m = 6$) to bad-metal ($m = 4$) to insulator ($m = 2$). In the extreme case, a pure and novel $J_{eff} = 1/2$ Mott insulating state is realized at the level of atomically thin monolayer $IrO_2$ with almost isolated $IrO_6$ octahedra.

We summarize the simulation results for the superlattices as well as for $IrO_2$ bulk ($m = \infty$), by presenting in Fig. 1 the phase diagram as a function of thickness $m$ and Coulomb parameter $U$. When including SOC, either without or with very tiny $U$, the ground states are essentially NM metallic for $IrO_2$/$TiO_2$ superlattices with $m \geq 2$ and for $IrO_2$ bulk. As shown in Fig. 1, crystal-field effects combined with SOC are insufficient to open a gap; however, electronic correlations have a crucial effect on the band structure. As the Coulomb interactions increase to a moderate critical value of $U_{c1}$, the superlattices ($m = 2, 4$ and $6$) as well as $IrO_2$ bulk ($m = \infty$) transform from NM metals to AFM metals; furthermore, a MIT is observed from AFM metal to AFM insulator at a higher critical value of $U_{c2}$ (for $m = 2$ $U_{c2} \approx 1$ eV and for $m = 4$ $U_{c2} \approx 2.5$ eV). For the extreme case of $m = 1$, the $(IrO_2)_1(TiO_2)_9$ superlattice - assumed to be ferromagnetic (FM), consistently with one Ir per unit-cell and with quite weak in-plane interactions between Ir in nearby unit cells (the energy difference between FM and AFM states being less than 0.5 meV/Ir) - shows an insulating behaviour even without any $U$.

According to previous literature [6, 36, 37], a good agreement between experiments (in terms of optical conductivity, electronic structure, magnetic properties) and DFT is achieved using $U = 2$ eV for Ir $5d$, the value we employed for most of the results to be discussed below. As shown in Figure S1, keeping the same strength of $U = 2$ eV, a MIT is observed as the IrO$_2$ layer decreases from bulk ($m = \infty$) to bilayer ($m = 2$) and monolayer ($m = 1$). For the bulk ($m = \infty$) and $m = 4, 6$, the electronic structures typically show a metallic behaviour, whereas for $m = 2$, the two pairs of bands around $E_F$ are split off by an insulating gap, as shown in Figure S2. For the monolayer case, bands around $E_F$ become very narrow and the insulating gap increases further, as detailed below.

The $m = 1$ superlattice is indeed particularly interesting and its electronic structure is presented in Fig. 2. Ir $t_{2g}$ bands are located around $E_F$, with O $2p$ bands at lower energy. Ir $t_{2g}$ states are separated by a remarkable gap from the empty Ti $3d$ states located at higher energy, implying that the TiO$_2$ substrate acts as an insulating blocking layer [38]. Consistently with low-spin states of Ir$^{4+}$ ($5d^5$) with partially filled $t_{2g}$ states [4, 8], GGA (Fig. 2 (a)) results in a metallic state with $t_{2g}$ states crossing $E_F$. As the Coulomb interactions increase to 2 eV, a spin-down $t_{2g}$ band is shifted above $E_F$ and a tiny insulating gap opens up (Fig. 2 (c)). Even without any Hubbard correction, SOC has a disruptive effect on the band structure (Fig. 2 (b)): two narrow Ir $t_{2g}$ bands around $E_F$ show a half-filled character and are split off by a tiny gap, clearly separated by a gap from the other four Ir $t_{2g}$ valence bands at lower energy, suggesting the narrow pair of bands around $E_F$ to be $J_{eff} = 1/2$ doublet states, and the other four Ir $t_{2g}$ valence bands to be $J_{eff} = 3/2$ quartet states (detailed discussion will be presented below and in the Supplemental Material [27]). Upon increasing $U$, the insulating gap is further increased, the half-filled $J_{eff} = 1/2$ doublet states being split further off with a remarkable insulating gap (Fig. 2 (d)). Our results are consistent with the proposed schematic energy diagrams for $5d^5$ Ir$^{4+}$ ions [7]. It should be noted that a small $U$ alone cannot account for the band gap within GGA + $U$, whereas a strong SOC is essential to trigger the Mott transition, leading to a half-filled $J_{eff} = 1/2$ Hubbard system.

The $J_{eff} = 1/2$ states can be further inspected by projected density of states (pDOS) and band decomposed charge density. For better clarity, an Ir-centered local coordinate system ($x, y, z$) defined in Fig. 2 (e) and (f) is employed, with $z$ along one of the Ir-O directions in the $ab$ plane, and $x, y$ approximately pointing towards O atoms. We checked the contribution of $d_{xy}$, $d_{yz}$, and $d_{zx}$ states and the partial charge density for the conduction band minimum (CBM) (the isolated spin-down band above $E_F$ in Fig. 2 (c)), and for the two isolated bands (CBM and valence band maximum (VBM)) around $E_F$ in Fig. 2 (d). Without SOC, the pDOS (Fig. 2 (c)) and anisotropic partial charge density (Fig. 2 (e)) confirm that bands around $E_F$ mainly derive from (local) $d_{yz}$ and $d_{zx}$ orbitals, whereas the $d_{xy}$ orbital in a lower energy range below $E_F$, this situation being similar to IrO$_2$ bulk [22]. In fact, we recall that in rutile structure, IrO$_6$ octahedra are distorted with slightly compressed local $z$ axis and largely distorted local $xy$ plane, breaking the degeneracy of the Ir $t_{2g}$ manifold, therefore split into a singlet ($d_{xy}$) and a quasi-degenerate doublet ($d_{xz}$ and $d_{yz}$) [24]. Returning to the monolayer upon inclusion of SOC, the pDOS (Fig. 2 (d)) indicates that the two bands around $E_F$ are derived from a mixture of the three Ir $t_{2g}$ orbitals with almost equal contributions from $d_{xy}$, $d_{yz}$, and $d_{zx}$ orbitals, "smoking gun" of the $J_{eff} = 1/2$ state [3, 7, 8]. Indeed, as clearly shown in Figure S3, the pDOS decomposed into the $J_{eff} = 1/2$ and 3/2 states, along with the isotropic isosurfaces of the partial charge density in Fig. 2 (f), further demonstrate the existence of a "pure" $J_{eff} = 1/2$ state [3].

Interestingly, the "effective-$J$" state shows a clear dependence on the connectivity of IrO$_6$ octahedra, the "purity" of the $J_{eff} = 1/2$ state being crucially affected by the local environment and structural distortions of IrO$_6$ octahedra [22,23]. As shown in the insets of Fig. 3, the connectivity of IrO$_6$ octahedra decreases upon decreasing the IrO$_2$ layer thickness $m$ (in $m = 4$ octahedra are corner-and-edge sharing as in bulk, in contrast corner-sharing in bilayer and isolated in monolayer). In turn, as shown by the pDOS in Fig. 3, this crucially tunes the bandwidth of $t_{2g}$ orbitals and, therefore, correlation effects and tendency towards Mott instability [39]. Moreover, without SOC (left column of Fig. 3), $t_{2g}$ states show a sizable splitting between the $d_{xy}$ and (almost degenerate) $d_{xz}$, $d_{yz}$ levels, the splitting decreasing upon reducing the thickness $m$ [21-24]. What happens upon including SOC? As shown in Fig. 3 (right column), the detailed components of the $t_{2g}$ states are basically unaltered with respect to those without SOC for bulk and for $m \geq 4$, whereas the contribution of the $t_{2g}$ states remarkably changes for thinner IrO$_2$ layers. In particular, while the connectivity reduces when decreasing thickness from bilayer to monolayer, the $d_{xy}$ contribution around $E_F$ rapidly increases. For bilayer, without SOC, the pDOS (Fig. 3) and anisotropic partial charge density (Figure S2 (e)) confirm that the CBM mainly derives from the two Ir $d_{yz}$ and $d_{zx}$ orbitals. In contrast, when including SOC, the pDOS (Fig. 3) as well as the band structure (Figure S2 (d)) indicate that the two pairs of bands around $E_F$ are derived from a mixture of the three Ir $t_{2g}$ orbitals with increasing contribution from $d_{xy}$ orbital, pointing towards a $J_{eff} = 1/2$ state. As for the effect of $U$, without $U$, two pairs of bands cross $E_F$ separated by a tiny gap from other Ir $t_{2g}$ valence bands within GGA + SOC (Figure S2 (b)), whereas MIT occurs within GGA + SOC + $U$. We may speculate that $t_{2g}$ states around $E_F$ originate from a mixing of $J_{eff} = 1/2$ doublet and $J_{eff} = 3/2$ quartet states, due to octahedral distortions. The existence of the $J_{eff} = 1/2$ state with lower purity compared with the ideal case is also confirmed by the partial charge density (Figure S2 (f) and (g)) showing a less isotropic character [3, 22, 40].

In summary, we have put forward rutile-based $IrO_2$/$TiO_2$ superlattices as new systems for emerging novel electronic states, where a metal-insulator transition can be tuned as a function of $IrO_2$ layer thickness $m$. The electronic structure in proximity to the Fermi level is found to be crucially affected by the connectivity of octahedra (in turn related to $m$): for isolated $IrO_6$ octahedra - closer to the ideal cubic crystal field with negligible inter-site effects - the strength of SOC competes with the noncubic crystal field splitting, resulting in a higher purity of the $J_{eff} = 1/2$ state. As the thickness is increased, the tetragonal crystal field splitting grows and overcomes SOC, in turn leading to a situation which is progressively farther from the "pure" $J_{eff} = 1/2$ state. We hope our results, based on density functional theory simulations, will stimulate experimental works aimed at verifying our predictions and will broaden the field of iridates-based spin-orbitronics.

**Acknowledgements**

We thank Carmine Autieri, Igor Di Marco and Fengren Fan for interesting discussions. This work was supported by the CARIPLO Foundation through the MAGISTER Project Rif. 2013-0726 and by IsC43 "C-MONAMI" Grant at Cineca Supercomputing Center. X. M. was sponsored by the China Scholarship Council (No. 201508420180), Natural Science Foundation of Hubei Province (No. 2014CFB439), Scientific and Technologic Research Program of DOE of Hubei Province (No. D20132902), Outstanding Young Science and Technology Innovation Team Program of Hubei Provincial Colleges and Universities (No. T201514).

**Figure Caption**

**Fig. 1** Phase diagram of $IrO_2/TiO_2$ superlattices and $IrO_2$ bulk, as a function of $IrO_2$ layer thickness $m$ and Hubbard parameter $U$. Coloured symbols show the points for which calculations have been performed; black diamond, green circle, blue positive-triangle and red inverted-triangle denote FM insulator (FMI), AFM insulator (AFMI), AFM metal (AFMM), and NM metal (NMM), respectively. SOC has been taken into account.

**Fig. 2** Electronic structure for $(IrO_2)_1(TiO_2)_9$ superlattices. Band structure and corresponding pDOS for Ir $t_{2g}$ states within (a) GGA, (b) GGA + SOC, (c) GGA + $U$, and (d) GGA + $U$ + SOC, where $U = 2$ eV. The partial charge density for the CBM in (c) within GGA + $U$ and (d) within GGA + $U$ + SOC are shown in (e) and (f), respectively. The insets in (a) and (b) shows the crystal field splitting and splittings of $t_{2g}$ by SOC, respectively. We use *xyz* for the local coordinates and *abc* for the global orientation (as shown in (e) and (f)).

**Fig. 3** Evolution of pDOS of Ir $t_{2g}$ states of $IrO_2/TiO_2$ superlattices ($m = 1, 2, 4$) and bulk ($m = \infty$) calculated with GGA + $U$ (left column) and GGA + $U$ + SOC (right column). Insets schematically show the connectivity of $IrO_6$ octahedra for the $IrO_2$ layer. Due to the structural symmetry, the pDOS is shown for only one type of Ir ion for $m = 1, 2$ and bulk, and for two types of Ir ion ($Ir_1/Ir_2$ denote the inner/interfacial Ir ion) for $m = 4$.

**Fig. 1**

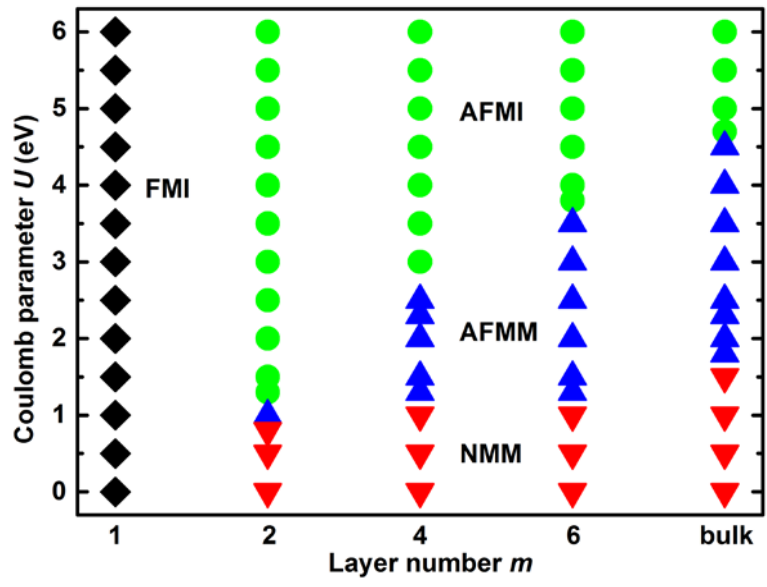

**Fig. 2**

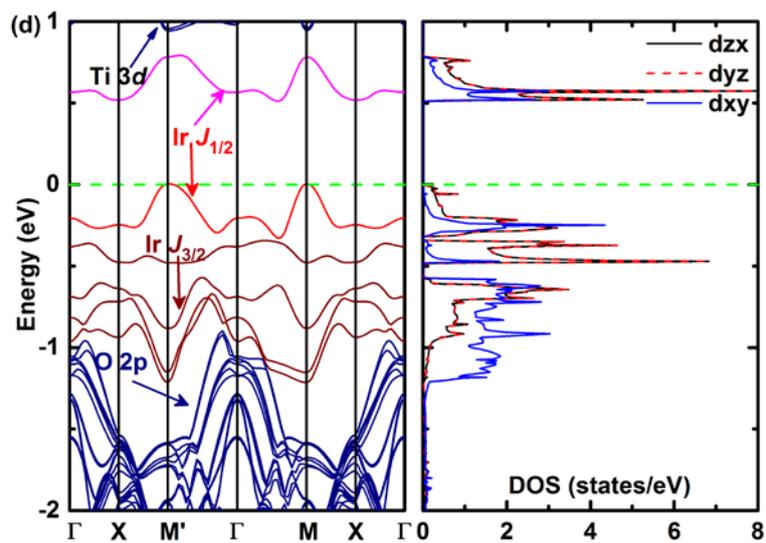

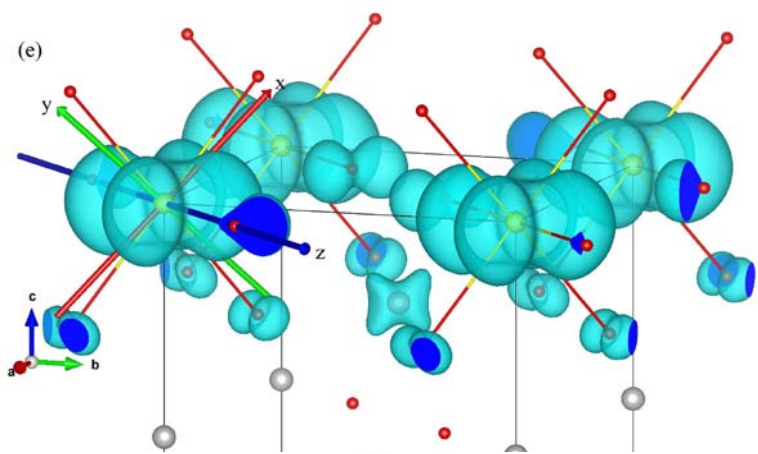

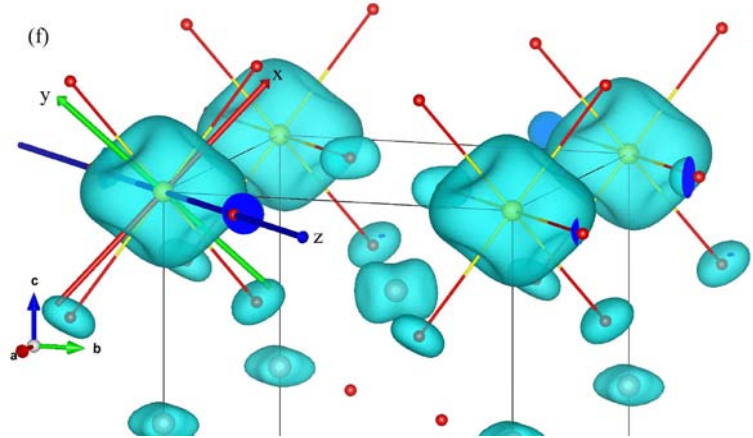

**Fig. 3**

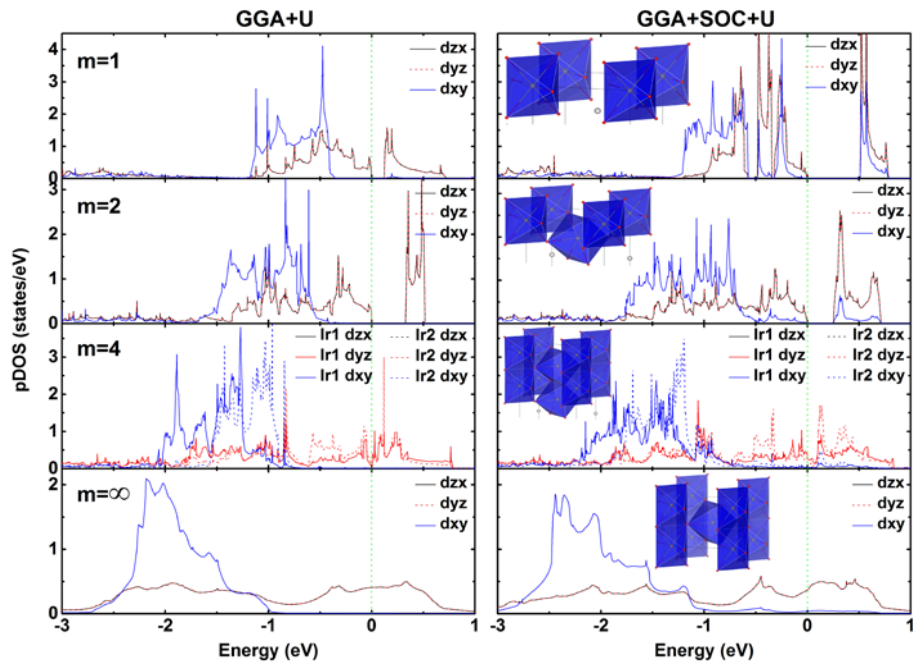

**Supplemental Material for**

**Metal-Insulator Transition and $J_{eff}$ =1/2 Spin-Orbit Insulating State in $IrO_2$/$TiO_2$ Superlattices**

Xing Ming,[1,2] Kunihiko Yamauchi,[3] Tamio Oguchi,[3] Silvia Picozzi[2]*

*1. College of Mathematics and Physics and Hubei Key Laboratory for Processing and Application of Catalytic Materials, Huanggang Normal University, Huanggang 438000, PR China*

*2. Consiglio Nazionale delle Ricerche CNR-SPIN, UOS L'Aquila, Sede Temporanea di Chieti, 66100 Chieti, Italy*

*3. ISIR-SANKEN, Osaka University, 8-1 Mihogaoka, Ibaraki, Osaka, 567-0047, Japan*

\* Email: silvia.picozzi@spin.cnr.it

## I. The Crystal Structure Model of the $IrO_2$/$TiO_2$ Superlattices

Both $IrO_2$ and $TiO_2$ crystallize in the rutile structure with two formula units per unit cell, where each metal atom M (M = Ir or Ti) lies in an $MO_6$ octahedral cage. The $MO_6$ octahedra share corners and edges to form a three-dimensional network in the space group $P_{42/mnm}$ (see detailed structure of $IrO_2$ bulk in Refs. 19 and 20). In the present study, ten layers of $TiO_2$ are considered as a matrix to embed ultrathin $IrO_2$ layers in the $(IrO_2)_m(TiO_2)_{10}$ superlattice ($m$ = 2, 4, 6), where the layer number $m$ refers to the number of $IrO_2$ formula units per unit cell. A special $(IrO_2)_1(TiO_2)_9$ superlattice ($m$ =1) is built by replacing one layer of $TiO_2$ in the ten layers $TiO_2$ matrix, in which isolated $IrO_6$ octahedra are neither corner, nor edge, nor face sharing with any other $IrO_6$ octahedra. The in-plane lattice constant is fixed to the optimized lattice constant of bulk $TiO_2$ (mimicking the heterostructure growth on a $TiO_2$ substrate), whereas the perpendicular lattice constant and all atomic internal positions are optimized.

## II. First-Principles Density-Functional Theory (DFT) Calculations

DFT calculations were performed using the Vienna Ab initio Simulation Package (VASP) code [29] within projector augmented wave (PAW) method.[30, 31] The generalized gradient approximation (GGA) exchange-correlation functional of Perdew-Burke-Ernzerhof functional revised for solids (PBEsol) was used for all calculations.[32] SOC was included using unconstrained noncollinear-magnetism settings. The rotationally invariant + U method was employed to account for correlations effects.[33] A $k$-point mesh of 11 × 11 × 1 (8× 8 × 12 for

IrO$_2$ bulk) and a cutoff energy of 520 eV have been used.

We have computed the electronic structure of IrO$_2$/TiO$_2$ superlattice and IrO$_2$ bulk with spin quantization axis in the *ab* plane (local [001] direction) and out-of plane (local [110] direction). Table S1 shows energy differences between the in-plane and out-of plane magnetizations. The electronic structure and energy show little difference for the calculated results with different quantization axis. With moderate on-site Coulomb interactions parameter $U \sim 2$ eV, the energy is a bit lower with the magnetic moment of Ir along the local [001] direction in the *ab* plane. So the main text just shown the results for quantization axis in the *ab* plane (along the local [001] direction).

Table S1 Calculated energies per Ir atom (meV) for different combinations of $U$ parameters with the quantization axis in the *ab* plane (local [001] direction) relative to out-of plane (local [110] direction). The energy for the quantization axis out of plane is taken as reference.

| U   | $m = 1$ | $m = 2$ | $m = 4$ | $m = 6$ | $m = \infty$ |
|-----|---------|---------|---------|---------|--------------|
| 0   | -7.37   | 0.00    | 0.00    | 0.00    | 0.01         |
| 0.5 | -3.41   | 0.00    | 0.00    | 0.02    | 0.01         |
| 1   | -1.81   | -0.04   | -0.04   | -0.19   | 0.03         |
| 1.5 | -0.68   | -10.66  | -2.88   | -1.13   | -0.08        |
| 2   | 0.16    | -9.44   | -6.99   | -0.63   | 0.13         |
| 2.5 | 0.85    | -8.96   | -11.55  | -4.94   | -1.10        |
| 3   | 1.41    | -8.84   | -18.22  | -13.71  | -4.71        |
| 3.5 | 1.85    | -8.83   | -17.22  | -6.86   | -12.30       |
| 4   | 2.23    | -8.82   | -16.28  | -18.34  | -19.04       |
| 4.5 | 2.54    | -8.74   | -15.31  | -17.33  | -20.48       |
| 5   | 2.84    | -8.59   | -14.30  | -16.24  | -19.36       |
| 5.5 | 3.13    | -8.37   | -13.23  | -15.03  | -18.06       |
| 6   | 3.52    | -8.08   | -12.10  | -13.67  | -16.43       |

### III. Spin-Orbit Eigenstates and Effective *J* State

In order to discuss the spin-orbit eigenstates and effective *J* states of the IrO$_2$/TiO$_2$ multilayer, here we clarify the relation between "correct" *J* states and "effective" *J* states by showing DFT-obtained electronic states.

#### A. Definition of Correct *J* State

When spin-orbit coupling is strong, angular momentum $L$ and spin momentum $S$ are no more good quantum numbers. Instead, we consider $J$ as a composition of $L+S$. The linear combination coefficients $|L,S,L_z,S_z|J,J_z\rangle$ are called Clebsh-Gordan coefficients, which are tabulated in literatures. Here we call it "correct" $J$ states.

### B. Definition of Effective $J$ State

The cubic crystal field states are described by real spherical harmonics as follows:

$$|3z^2-r^2\rangle = Y_2^0$$

$$|x^2-y^2\rangle = \frac{1}{\sqrt{2}}\left(Y_2^{-2}+Y_2^2\right)$$

$$|xy\rangle = \frac{i}{\sqrt{2}}\left(Y_2^{-2}-Y_2^2\right)$$

$$|yz\rangle = \frac{i}{\sqrt{2}}\left(Y_2^{-1}+Y_2^1\right)$$

$$|zx\rangle = \frac{1}{\sqrt{2}}\left(Y_2^{-1}-Y_2^1\right)|zx\rangle$$

The first two are $e_g$ states and the last three are $t_{2g}$ states. By neglecting the $e_g$ states and operating angular operators $L_z$, $L_+$, $L_-$, the $t_{2g}$ states are transformed as in Table S2.

TABLE S2: Transformation of the $t_{2g}$ states under $L$ operator

|       | $|xy\rangle$ | $|yz\rangle$ | $|zx\rangle$ |
|-------|---|---|---|
| $L_z$ | 0 | $-i|zx\rangle$ | $i|yz\rangle$ |
| $L_+$ | $|yz\rangle+i|zx\rangle$ | $-|xy\rangle$ | $-i|xy\rangle$ |
| $L_-$ | $-|yz\rangle+i|zx\rangle$ | $|xy\rangle$ | $-i|xy\rangle$ |

The transformation of $t_{2g}$ states shows equivalence to the correct cubic harmonics for $L=1$ $(xy \rightarrow -pz, yz \rightarrow -px, zx \rightarrow -py)$, so that we define "effective" $L$ states for $L^{eff}=1$ ($|L^{eff}=1, L_z^{eff}=1,0,-1\rangle$) given by

$$|1,1\rangle = \frac{1}{\sqrt{2}}(|yz\rangle + i|zx\rangle)$$

$$|1,0\rangle = -|xy\rangle$$

$$|1,-1\rangle = \frac{1}{\sqrt{2}}(-|yz\rangle + i|zx\rangle)$$

By using the above definition, we now compose $L^{eff}=1$ and $S=1/2$ to make $J^{eff}=3/2$ states through Clebsh-Gordon coefficients as $|J^{eff}, J_z^{eff}\rangle = \sum C.G. |L^{eff}, S\rangle$.

For $|J^{eff}=3/2\rangle$, we end up with

$$|3/2, 3/2\rangle = |1,\uparrow\rangle = \frac{1}{\sqrt{2}}(|yz,\uparrow\rangle + i|zx,\uparrow\rangle)$$

$$|3/2, 1/2\rangle = \frac{1}{\sqrt{3}}|1,\downarrow\rangle + \frac{\sqrt{2}}{\sqrt{3}}(|0,\uparrow\rangle) = \frac{1}{\sqrt{6}}(|yz,\downarrow\rangle + i|zx,\downarrow\rangle - 2|xy,\uparrow\rangle)$$

$$|3/2, -1/2\rangle = \frac{1}{\sqrt{3}}|-1,\uparrow\rangle + \frac{\sqrt{2}}{\sqrt{3}}(|0,\downarrow\rangle) = \frac{1}{\sqrt{6}}(-|yz,\uparrow\rangle + i|zx,\uparrow\rangle - 2|xy,\downarrow\rangle)$$

$$|3/2, -3/2\rangle = |-1,\downarrow\rangle = \frac{1}{\sqrt{2}}(-|yz,\downarrow\rangle + i|zx,\downarrow\rangle)$$

For $|J^{eff}=1/2\rangle$, we end up with

$$|1/2, 1/2\rangle = \frac{\sqrt{2}}{\sqrt{3}}|1,\downarrow\rangle - \frac{1}{\sqrt{3}}(|0,\uparrow\rangle) = \frac{1}{\sqrt{3}}(|yz,\downarrow\rangle + i|zx,\downarrow\rangle + |xy,\uparrow\rangle)$$

$$|1/2, -1/2\rangle = -\frac{\sqrt{2}}{\sqrt{3}}|-1,\uparrow\rangle + \frac{1}{\sqrt{3}}(|0,\downarrow\rangle) = \frac{1}{\sqrt{3}}(|yz,\uparrow\rangle - i|zx,\uparrow\rangle - |xy,\downarrow\rangle)$$

The last two states are $J^{eff}=1/2$ states discussed in the iridates.

### C. Projected Density of States (DOS) of Effective *J* State

In order to obtain *J*-projected density of states (DOS) of monolayer $IrO_2$ embedded in $TiO_2$ multilayers, $(IrO_2)_1/(TiO_2)_9$, we performed a DFT calculation. Since the *J* projection scheme is not implemented in the VASP code, we used the HiLAPW code [34] within GGA-PBE+*U* approximation. *U* and *J* values are set as 2.0 eV and 0.2 eV, respectively. The Kohn-Sham equations are solved self-consistently by using the all-electron scalar-relativistic full-potential linearized augmented plane-wave (FLAPW) method.[35, 36] The crystal structure setting is the same as the one in the main text. The $k$-space integrations are done with the improved tetrahedron method [37] with $8 \times 8 \times 2$ $k$ mesh. Spin-orbit coupling is treated as the second-variation step [38] in the self-consistent loop with spin quantization axes parallel to the Ir-O bond direction in the IrO$_2$ *ab* plane (local [001] direction).

Figure S3 summarizes the calculated DOS of (IrO$_2$)$_1$/(TiO$_2$)$_9$ with different projections. Figure S3 (a) and (b) show that the mixing of $t_{2g}$ states forms the $|J=3/2\rangle$ and $|J=5/2\rangle$ states. The upper and lower Mott bands are assigned as $|J, J_z\rangle = |5/2, -3/2\rangle$ and $|5/2, 3/2\rangle$ states, respectively. By using projection as explained above, the effective *J* states are shown in Figure S3 (c). The upper and lower Mott bands are now $|J^{eff}, J_z^{eff}\rangle = |1/2, 1/2\rangle$ and $|1/2, -1/2\rangle$ states, respectively, which confirms the monolayer IrO$_2$ as the effective $J^{eff} = 1/2$ Mott insulator.

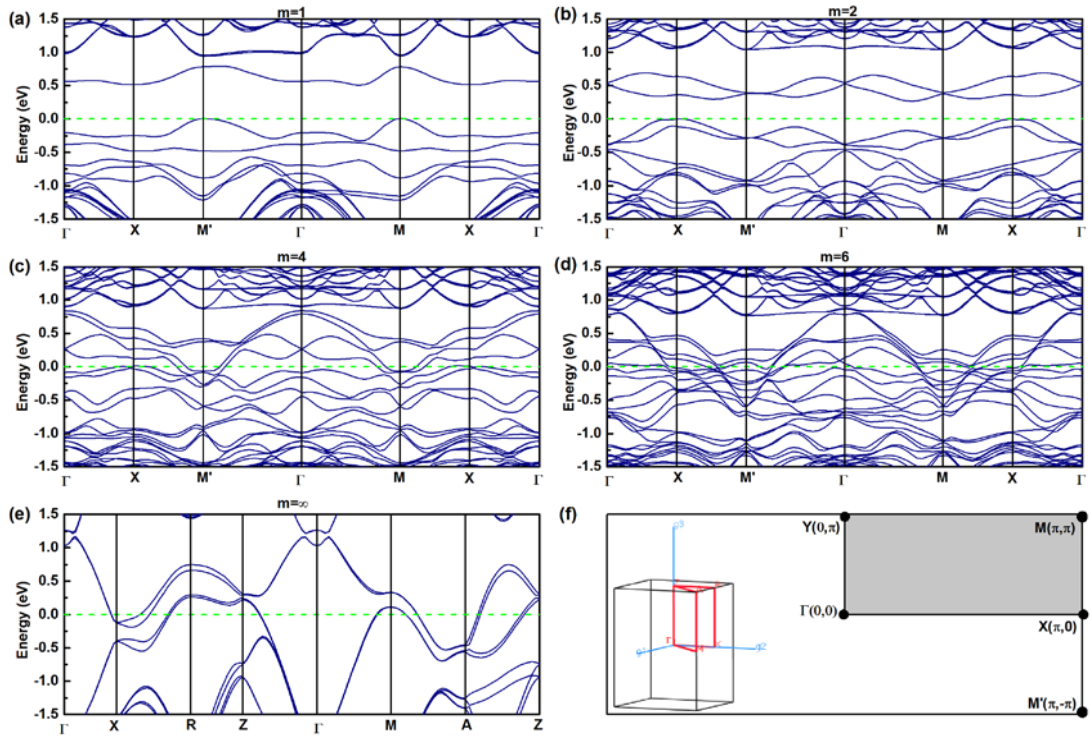

Figure S1 Detailed band structure of IrO$_2$/TiO$_2$ superlattices and the IrO$_2$ bulk ($m = \infty$) calculated with GGA + SOC + $U$ ($U = 2$ eV) for (a) $m = 1$, (b) $m = 2$, (c) $m = 4$, (d) $m = 6$, (e) $m = \infty$ and (f) the Brillouin zone for the superlattice, the inset in (f) is the Brillouin zone of the IrO$_2$ bulk.

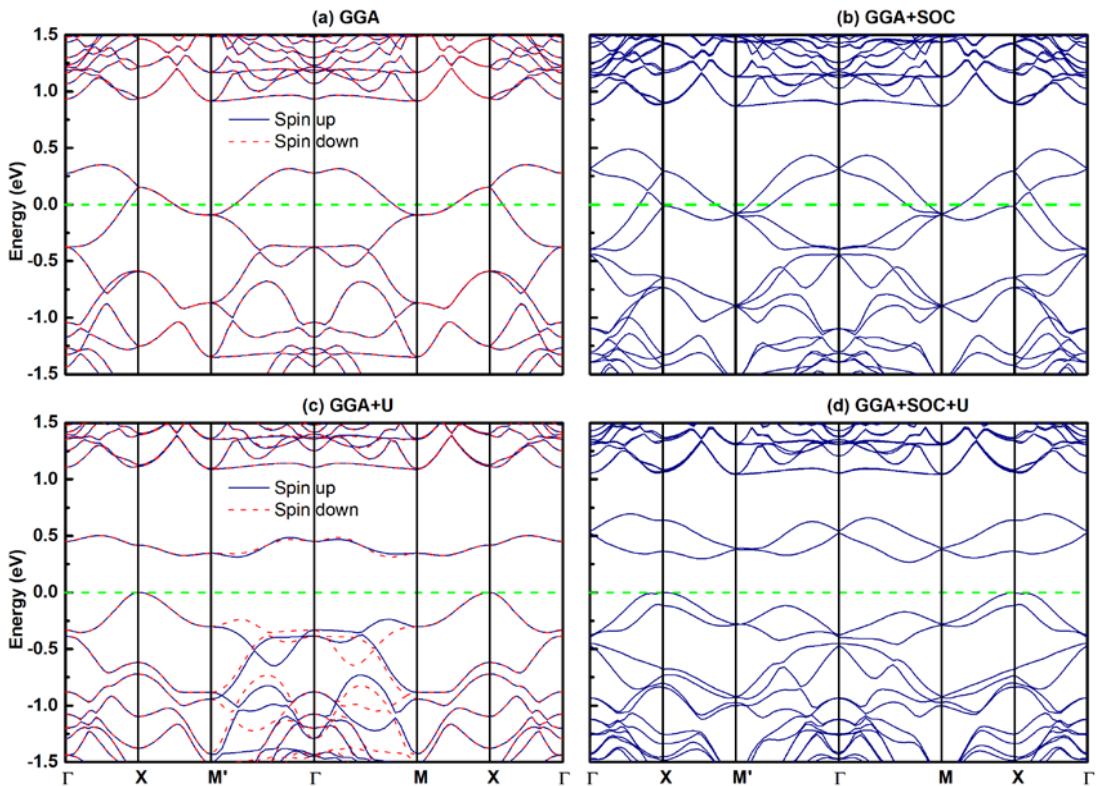

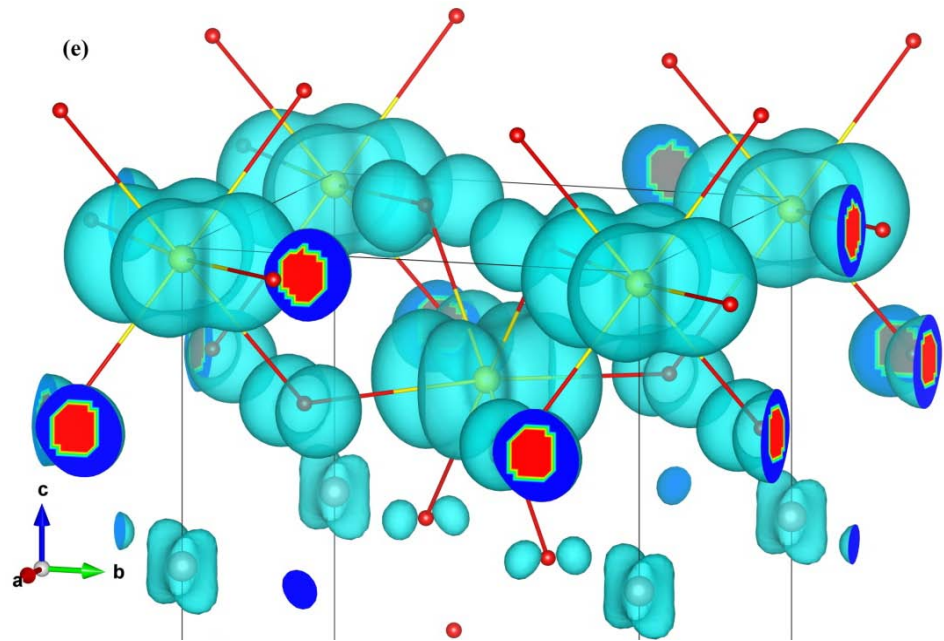

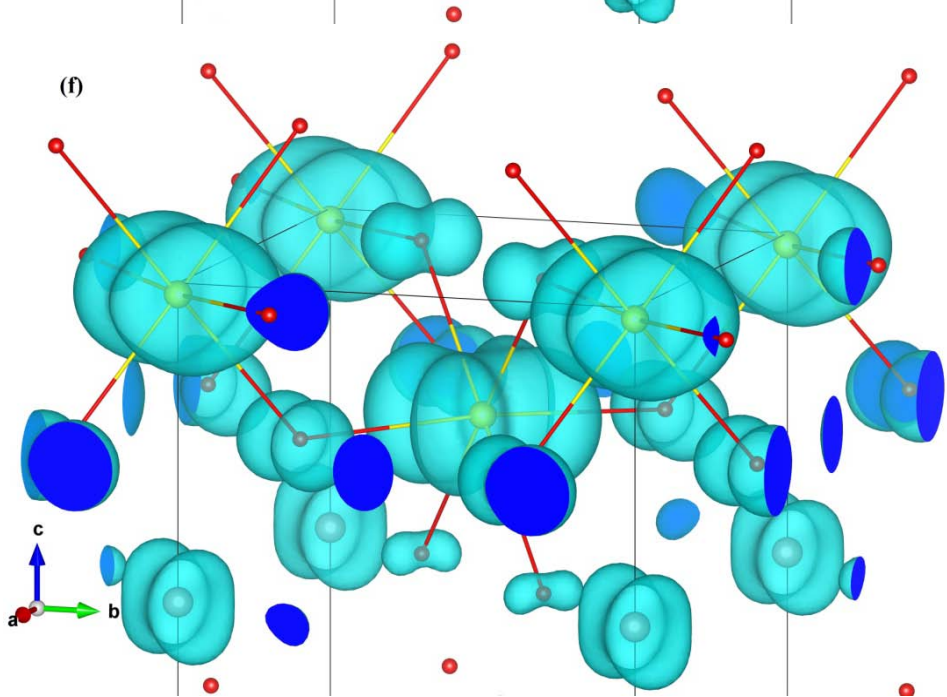

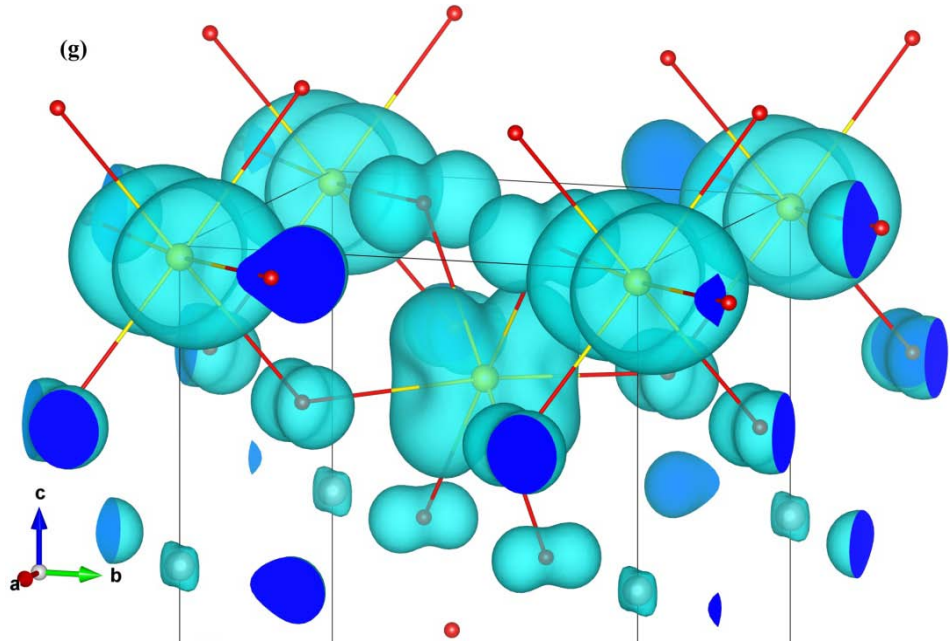

Figure S2 Detailed electronic structure of $(IrO_2)_2(TiO_2)_{10}$ superlattices: The band structure calculated with (a) GGA, (b)and GGA + SOC, (c) GGA + $U$, (d) GGA + SOC + $U$ and the partial charge density for the CBM as shown in (c), and the CBM (pair of conduction band) and valence band maximum (VBM) (pair of valence band) in (d) are shown in (e), (f) and (g), respectively.

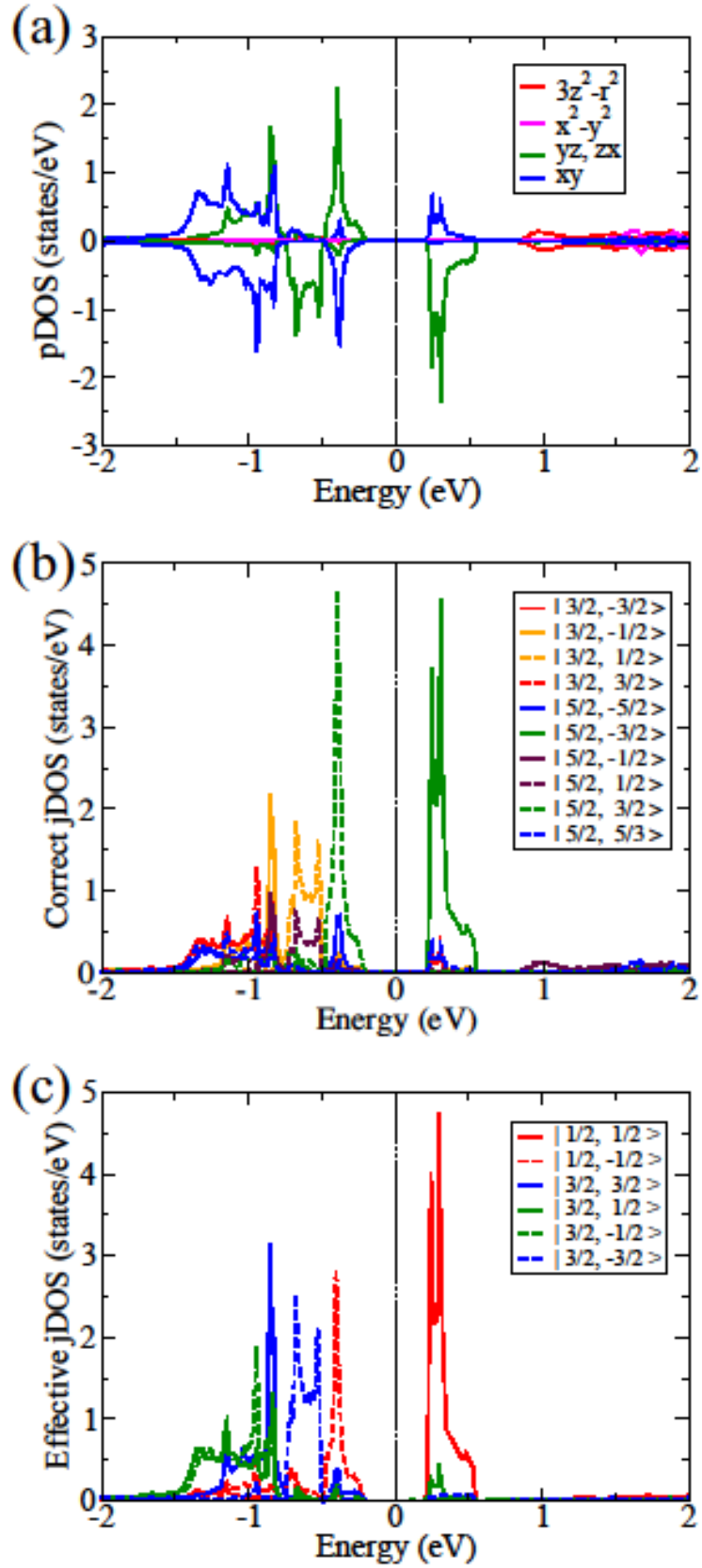

Figure S3 DOS projected onto (a) $L = 2$ cubic harmonics with spin states, (b) correct $J$ ($|J, J_z\rangle$) states, and (c) effective $J$ ($|J^{eff}, J_z^{eff}\rangle$) states.